\documentclass[reprint, amsmath,amssymb,aps]{revtex4-2}

\usepackage{graphicx}
\usepackage{dcolumn}
\usepackage{bm}
\usepackage{amsmath}
\usepackage[usenames,dvipsnames]{xcolor}
\usepackage{mathtools}
\usepackage[colorlinks=true,citecolor=blue,linkcolor=blue]{hyperref}
\usepackage[caption=false]{subfig}
\usepackage{bbm}
\usepackage{xcolor}
\usepackage[color]{xy}  
\usepackage[braket]{qcircuit}

\UseCrayolaColors
\makeatletter
\newcommand{\outputgroupv}[7]{%
	\POS"#1,#3"."#2,#3"!C+<#4,0em>*{\Bigg\}}%
	\POS"#1,#3"."#2,#3"!C+<#5,#6>*{#7}%
}
\makeatother
\usepackage{tikz}
\newcommand{\metersymbol}{%
	\begin{tikzpicture}[baseline=-0.5ex,scale=0.42]
		\draw (-0.8,-0.45) arc[start angle=180,end angle=0,radius=0.8];
		\draw (-0.8,-0.45) -- (0.8,-0.45);
		\draw[->] (0,-0.45) -- (0.67,0.42);
	\end{tikzpicture}%
}
\usepackage{comment}
\usepackage{hyperref}
\usepackage{upgreek}
\usepackage{esint}
\usepackage{soul}
\usepackage[normalem]{ ulem } 
\usepackage{floatrow}
\usepackage{verbatim}
\usepackage{braket}
\usepackage{algorithm}
\usepackage{algpseudocode}
\usepackage{float}
\usepackage{array}
\usepackage{placeins}
\newcommand{\PreserveBackslash}[1]{\let\temp=\\#1\let\\=\temp}
\newcolumntype{C}[1]{>{\PreserveBackslash\centering}p{#1}}

\newcommand{\rom}[1]{\uppercase\expandafter{\romannumeral #1\relax}}

\begin{document}

\title{Solving wave propagation problems via geometric quantum state preparation on dispersion manifolds}

\author{Yakov Solomons}\email{yakov.solomons@quantum-art.tech}
\author{Lee Peleg}
\author{Netanel Barel}
\author{Jonathan Nemirovsky}
\author{Amit Ben-Kish}
\author{Yotam Shapira}

\affiliation{Quantum Art, Ness Ziona 7403682, Israel}

\begin{abstract}
We present a quantum algorithm for solving partial differential equations through a linear-system formulation, focusing on wave propagation problems described by the discretized Helmholtz equation in frequency domain. Although quantum linear-system solvers offer exponential compression of the system degrees of freedom, their runtime complexity is generally governed by the condition number of the discretized operator. Exploiting the analytic structure of the differential operator can provide an alternative to explicit matrix inversion, as illustrated for the screened Poisson equation through an explicit quantum circuit. For the Helmholtz equation, however, the inverse operator becomes singular on the dispersion surface $k^2=\omega^2/c^2$, rendering direct Fourier-space state-preparation methods exponentially inefficient. We address this challenge by directly preparing quantum states supported on the resonant manifold and encoding source locations through Fourier phases. The resulting algorithm eliminates the exponentially large overhead associated with post-selection on the resonant manifold, yielding a success probability that depends linearly on the number of sources and is independent of the computational domain size. More generally, our approach applies to hyperbolic differential equations whose Fourier-space solutions possess singular support on dispersion manifolds, recasting their solution as a problem of geometric quantum state preparation.

\end{abstract}

\maketitle

\section{Introduction}

The wave equation is a fundamental partial differential equation (PDE) describing the propagation of waves in a wide variety of physical systems, including acoustics, electromagnetism, fluid dynamics, and elastic media. The development of efficient methods for solving the wave equation has long been an important topic in computational science. In recent years, quantum computing has emerged as a promising framework for the simulation of physical systems, leading to the development of several quantum algorithms for solving the wave equation \cite{costa2019quantum,suau2021practical,sato2024hamiltonian,tezuka2026quantum,shringi2026structure}. Quantum algorithms naturally offer exponential advantage, as $2^n$ field points can be encoded on only $n$ qubits.

Quantum algorithms for solving the wave equation can be broadly divided into two categories. The first is based on Hamiltonian simulation, in which the wave dynamics is encoded into a Schrödinger-like evolution generated by an effective Hamiltonian \cite{costa2019quantum,suau2021practical,sato2024hamiltonian,tezuka2026quantum}. Closely related formulations have been developed in the context of coupled classical oscillators \cite{babbush2023exponential,luangsirapornchai2025practical}. This approach offers an exponential compression of the spatial degrees of freedom, enabling the simulation of very large domains. However, its complexity scales linearly with the simulated evolution time, which itself grows exponentially with the number of qubits, since the relevant evolution time is typically set by the time required for the wave to propagate across the entire simulated domain.

The second approach reformulates the discretized wave equation as a system of linear equations and employs a Quantum Linear Systems Solver (QLSS) to obtain the solution, e.g, the well known HHL algorithm \cite{harrow2009quantum}. This paradigm is widely used in quantum PDE solvers \cite{childs2021high,berry2014high,cao2013quantum,wang2020quantum,wang2024quantum,bagherimehrab2023fast}. In this framework, both space and time can be discretized and encoded into the quantum state, with each basis state corresponding to a point in space-time. The wave equation is then represented as a linear system whose solution contains the field values at all spatial and temporal grid points simultaneously, suggesting the possibility of achieving exponential quantum speedups in both space and time.

Alternatively, one may consider a frequency-domain formulation of the problem. In many applications, the sources are monochromatic or concentrated within a narrow frequency band, making such a formulation particularly relevant. In this setting, the field $u(\omega,\mathbf{r})$ satisfies the \emph{Helmholtz equation} \cite{gu2025quantum}
\begin{equation}
	\left(\frac{\omega^2}{c^2} + \nabla^2\right)u(\omega,\mathbf{r})
	=
	f(\omega,\mathbf{r}),
	\label{helmholtz}
\end{equation}
where $\omega$ is the temporal frequency, $\mathbf{r}$ denotes the spatial coordinates, $c$ is the wave velocity, and $f(\omega,\mathbf{r})$ represents the source term. As in the time-domain formulation, this also raises the possibility of exponential quantum speedups, as solving the discretized Helmholtz equation with a QLSS can yield a quantum representation of the steady-state field with complexity scaling only polylogarithmically with the domain size.

However, this apparent exponential quantum advantage faces an important obstacle. In many cases, a matrix $A$ arising from the discretization of partial differential equation will exhibit a condition number $\kappa(A)=\|A\|\,\|A^{-1}\|$ that grows with the domain size \cite{tong2021fast}. This presents a fundamental challenge for QLSSs. Although the complexity of QLSSs depends directly only polylogarithmically on the dimensions of $A$ (as in the HHL algorithm \cite{harrow2009quantum}), it also scales at least linearly with the condition number, and this scaling is optimal \cite{tong2021fast}. Thus, any growth of the condition number with the domain size is directly reflected in the algorithmic complexity. Consequently, the expected exponential speedup with respect to domain size is generally lost.

Ref.~\cite{tong2021fast} proposed a strategy to circumvent the condition-number dependence of generic QLSS algorithms by exploiting the known structure of the differential operator. In the Fourier basis, the operator is diagonal, $\frac{\omega^2}{c^2}+\nabla^2\;\rightarrow\;\frac{\omega^2}{c^2}-k^2$, allowing the solution state to be prepared directly using quantum arithmetic, without explicitly inverting the discretized matrix. For the wave equation, however, the inverse operator $(\omega^2/c^2-k^2)^{-1}$ is singular on the resonant manifold
\begin{equation}
	k^2=\frac{\omega^2}{c^2},
\end{equation}
so the solution is concentrated near resonance. Preparing the full Fourier state via reversible quantum arithmetic as in Ref.~\cite{tong2021fast} yields a post-selection success probability set by the ratio between the resonant-surface and the total Fourier-space volume, requiring a number of repetitions that grows polynomially with the domain size and thus eliminates the exponential compression.

To overcome this difficulty, we propose a different strategy that prepares the resonant manifold directly. This construction is motivated by the observation that a point source has a particularly simple Fourier-space representation: an equal superposition of Fourier modes with phases determined by the source coordinates. Accordingly, the algorithm first prepares a uniform superposition over the resonant manifold and then imprints the corresponding Fourier phases. The resulting state directly represents the propagated wave in Fourier space. Although the algorithm still requires post-selection, its success probability depends only on the number of sources and is independent of the size of the computational domain. Since many practical applications involve only a small number of emitters, this introduces only a constant overhead, thereby preserving polylogarithmic scaling with the size of the computational domain.

More broadly, the same principle can be applied to other differential equations whose spectral-space solutions are supported on lower-dimensional manifolds. Examples include wave propagation in anisotropic or elastic media, where the relevant manifolds are determined by the corresponding dispersion relations. In such cases, for boundary conditions compatible with diagonalization in a spectral basis, solving the differential equation is transformed into preparing the corresponding geometric structure in spectral space and subsequently encoding the appropriate phase information. This shifts the computational challenge from matrix inversion to geometric state preparation suggesting a new approach of efficient quantum algorithms for wave propagation and related physical systems.

\section{Quantum Linear Systems Solvers for the Helmholtz Equation}

\subsection{Problem formulation}

We begin by discretizing the spatial domain to obtain a linear algebraic system. Starting with one spatial dimension, consider a uniform $N$ point grid with spacing $h$ and periodic boundary conditions. The Laplacian operator is approximated using the finite-difference expression

\begin{equation}
	\nabla^2 u(x_j) \approx \frac{u_{j+1} - 2u_j + u_{j-1}}{h^2}.
\end{equation}

Substituting this into the Helmholtz equation \eqref{helmholtz} yields

\begin{equation}
	\left(\frac{\omega^2}{c^2} - \frac{2}{h^2}\right)u_j + \frac{1}{h^2}u_{j+1} + \frac{1}{h^2}u_{j-1} = f_j.
\end{equation}

The discretized problem therefore takes the form of a linear system

\begin{equation}
	A\mathbf{u} = \mathbf{b},
	\label{linear_system}
\end{equation}

where

\begin{equation}
	\mathbf{u} = (u_0,\dots,u_{N-1})^T,
	\qquad
	\mathbf{b} = (f_0,\dots,f_{N-1})^T.
\end{equation}

The matrix $A$ is a circulant tridiagonal matrix,

\begin{equation}
	A =
	\begin{pmatrix}
		d & \alpha & 0 & \cdots & \alpha \\
		\alpha & d & \alpha & \cdots & 0 \\
		0 & \alpha & d & \cdots & 0 \\
		\vdots & \vdots & \vdots & \ddots & \vdots \\
		\alpha & 0 & 0 & \cdots & d
	\end{pmatrix}.
\end{equation}

with coefficients $d = \frac{\omega^2}{c^2} - \frac{2}{h^2}$, $\alpha = \frac{1}{h^2}$. The nonzero corner entries reflect the imposed periodic boundary conditions, coupling the first and last grid points.

In two spatial dimensions, the Laplacian operator becomes

\begin{equation}
	\nabla^2 u = \frac{\partial^2 u}{\partial x^2} + \frac{\partial^2 u}{\partial y^2}.
\end{equation}

Discretizing on a rectangular grid, leads to the standard five-point stencil,

\begin{equation}
	\nabla^2 u_{i,j} \approx \frac{u_{i+1,j} + u_{i-1,j} + u_{i,j+1} + u_{i,j-1} - 4u_{i,j}}{h^2}.
\end{equation}

The resulting discretized Helmholtz equation again produces a linear system of the form of Eq. \eqref{linear_system}, where $A$ is now a sparse matrix whose nonzero entries couple each grid point to its nearest neighbors on the two-dimensional lattice. This structure is characteristic of local differential operators and naturally follows the five-point stencil connectivity pattern. Throughout the remainder of this work we restrict our discussion to the two-dimensional case with equally spaced grid; the extension to three dimensions is straightforward.

QLSS aims to prepare a quantum state proportional to the solution of the linear system \cite{harrow2009quantum,tong2021fast,childs2017quantum,morales2411quantum}. Given a suitable encoding of the right-hand side vector $\mathbf{b}$ as a quantum state $\ket{b}$, the goal is to generate the state

\begin{equation}
	\ket{u} \propto A^{-1}\ket{b},
\end{equation}

which encodes the solution vector $\mathbf{u}$ up to magnitude factor. The central computational task is therefore the implementation of the matrix inverse $A^{-1}$ on a quantum computer.

The first quantum algorithm for this task was the celebrated Harrow-Hassidim-Lloyd (HHL) algorithm~\cite{harrow2009quantum}, which demonstrated that sparse linear systems can be solved with complexity that depends only polylogarithmically on the system size. The complexity of HHL scales as $\mathcal{O}\left(\kappa(A)^2,\mathrm{polylog}(N,1/\epsilon)\right)$, where $\kappa(A)$ is the condition number of the matrix, $\epsilon$ is the desired accuracy, and $N$ is the matrix dimension. Subsequent developments substantially improved this dependence, and the best known QLSS now achieve the optimal linear scaling with the condition number~\cite{costa2022optimal}. Among these developments, Quantum Singular Value Transformation (QSVT) has emerged as a powerful and general framework for implementing matrix functions on quantum computers~\cite{gilyen2019quantum,martyn2021grand,dong2024feedforward,motlagh2024generalized}. The key idea is to encode the matrix $A$ into a larger unitary operator, known as a block encoding \cite{chakraborty2018power,clader2023quantum,sunderhauf2024block}, and then interleave applications of this unitary with phase rotations to implement a polynomial transformation of its singular values. By approximating the function $f(x)=1/x$ \cite{pyqsp,sunderhauf2025matrix}, QSVT implements matrix inversion and thereby produces a state proportional to $A^{-1}\ket{b}$. QSVT-based QLSS achieve complexity

\begin{equation}
	\mathcal{O}\big(\kappa(A),\mathrm{polylog}(N,1/\epsilon)\big),
	\label{complexity}
\end{equation}

which is known to be optimal with respect to the dependence on the condition number~\cite{tong2021fast}. 

\subsection{The condition number problem}

The polylogarithmic dependence of QLSSs on the system size suggests an exponential quantum advantage over classical linear system solvers. However, for matrices arising from the discretization of differential equations, the condition number typically grows with domain size. Since the complexity of any QLSS necessarily scales at least linearly with the condition number, this growth is directly reflected in the algorithmic complexity and can eliminate the apparent exponential speedup.

Consider our second-order differential operator, \eqref{helmholtz} in $d$ spatial dimensions, discretized on a grid containing $N$ points. Following an analysis similar to that of Ref.~\cite{tong2021fast}, one finds

\begin{equation}
	\|A^{-1}\|=\mathcal{O}(1),
	\qquad
	\|A\|=\mathcal{O}(h^{-2}),
\end{equation}

where $h\sim N^{-1/d}$ is the grid spacing in $d$ spatial dimensions. Consequently,

\begin{equation}
	\kappa(A)
	=
	\|A\|\,\|A^{-1}\|
	=
	\mathcal{O}(N^{2/d}).
\end{equation}

Therefore, the complexity \eqref{complexity} for optimal QLSSs becomes

\begin{equation}
	\mathcal{O}\big(N^{2/d},\mathrm{polylog}(N,1/\epsilon)\big),
\end{equation}

eliminating the exponential advantage of the quantum representation.

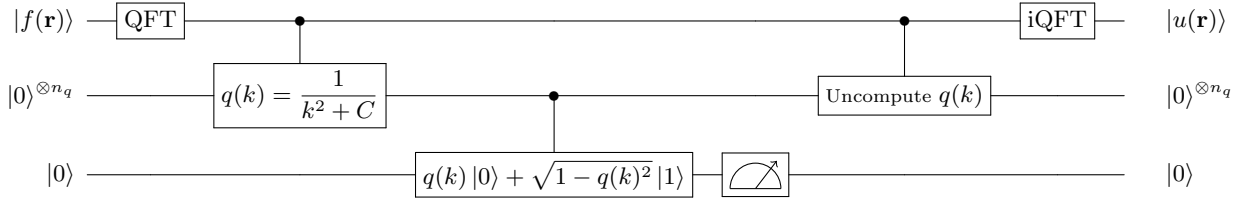
\begin{figure*}[t]
	\centering
\[
\Qcircuit @C=1.2em @R=1.0em {
	\lstick{\ket{f(\mathbf r)}}
	& \gate{\mathrm{QFT}}
	& \ctrl{1}
	& \qw
	& \qw
	& \ctrl{1}
	& \gate{\mathrm{iQFT}}
	& \qw
	& \rstick{\ket{u(\mathbf r)}}
	\\
	\lstick{\ket{0}^{\otimes n_q}}
	& \qw
	& \gate{\displaystyle q(k)=\frac{1}{k^{2}+C}}
	& \ctrl{1}
	& \qw
	& \gate{\text{\scriptsize Uncompute }q(k)}
	& \qw
	& \qw
	& \rstick{\ket{0}^{\otimes n_q}}
	\\
	\lstick{\ket{0}}
	& \qw
	& \qw
	& \gate{\displaystyle
		q(k)\ket{0}+\sqrt{1-q(k)^2}\ket{1}}
	& \gate{\metersymbol}
	& \qw
	& \qw
	& \qw
	& \rstick{\ket{0}}
}
\]
\caption{
	Quantum implementation of the inverse operator for the screened Poisson equation in Fourier space. 
	The Quantum Fourier Transform (QFT) maps the state to the momentum basis, where the differential operator becomes diagonal, this corresponds to the $U$ operation in the main text . 
	In this basis, inversion reduces to evaluating the scalar function
	$
	q(k)=1/(k^2+C)
	$, \eqref{q_k}.
	This value is computed by quantum arithmetic and then encoded into an auxiliary qubit amplitude through a controlled rotation, \eqref{aux_rotation}, producing the state
	$
	q(k)\ket{0}+\sqrt{1-q(k)^2}\ket{1}
	$.
	Finally, the inverse Quantum Fourier Transform (iQFT) returns the solution to the original basis.
}
\label{fig:poisson_qft_inverse}
\end{figure*}

\subsection{The basis transformation and quantum arithmetic algorithm}

A natural way to overcome this limitation is to exploit the structure of the matrix $A$. As discussed in~\cite{tong2021fast}, matrices arising from discretized differential equations often admit a simple representation in a suitable basis, such as the Fourier basis. Under a unitary transformation $U$, the matrix may become diagonal, $\tilde A = U^\dagger A U = \mathrm{diag}(\lambda_1,\ldots,\lambda_N)$, with analytically known eigenvalues $\lambda_j$. The inverse is then simply $\tilde A^{-1}=\mathrm{diag}(1/\lambda_1,\ldots,1/\lambda_N)$, so that $A^{-1}=U\tilde A^{-1}U^\dagger$.

Implementing $A^{-1}$ therefore reduces to applying $U^\dagger$, evaluating the known function $1/\lambda_j$ for each basis state $\ket{j}$ using quantum arithmetic, and applying $U$. In this approach, the inverse is constructed directly from its analytic form rather than through a general QLSS. Consequently, the complexity is determined by the cost of the arithmetic operations rather than by the condition number of the matrix.

As a simple example, consider the screened Poisson equation
\begin{equation}
	\left(C-\nabla^2\right)u(\mathbf r)=f(\mathbf r),
\end{equation}
where $C>0$ is a constant. This equation has almost the same form as the wave equation~\eqref{helmholtz}, differing only in the sign in front of the Laplacian operator $\nabla^2$. In Fourier space, the equation is diagonal, with
$A_F=\mathrm{diag}(C+k^2)$, $k=\sqrt{k_x^2+k_y^2}$, hence $A_F^{-1}=\mathrm{diag}\!\left(1/(C+k^2)\right)$.
Thus, the inverse can be implemented by evaluating
\begin{equation}
	q(k)=\frac{1}{C+k^2}
	\label{q_k}
\end{equation}
using quantum arithmetic. Specifically, a quantum reciprocal operation controlled by the register $\ket{k}$ computes $q(k)$ and stores it in an auxiliary register of $n_q$ qubits. The value is encoded as a binary fraction, with $n_q$ chosen according to the desired precision. For example, $n_q=10$ provides a resolution of $2^{-10}=1/1024\approx10^{-3}$ independent on the domain size (since the precision of $q(k)$ is decoupled from the resolution of the $k$-grid). The value $q(k)$ is then encoded into an auxiliary amplitude through a circuit that implements the transformation (e.g., \cite{sanders2019black})
\begin{equation}
	\ket{k}\ket{0}
	\;\longrightarrow\;
	\ket{k}
	\left(
	q(k)\ket{0}
	+
	\sqrt{1-q(k)^2}\ket{1}
	\right),
	\label{aux_rotation}
\end{equation}
followed by post-selection of the auxiliary qubit in the state $\ket{0}$. A circuit diagram of this algorithm is shown in Fig. \ref{fig:poisson_qft_inverse}. This example implements the idea introduced in Ref.~\cite{tong2021fast} of replacing a generic QLSS with direct reciprocal computation.

\subsection{The wave-equation singularity and the post-selection problem}

While the same quantum-arithmetic framework can, in principle, be applied to the wave equation, which is the focus of this work, the hyperbolic nature of the problem introduces an additional complication as the inverse operator develops singularities on the dispersion surface. After Fourier transforming, the wave equation \eqref{helmholtz} takes the form
\begin{equation}
	\left(\omega^2/c^2-k^2\right)\tilde{u}(k) = \tilde{f}(k),
\end{equation}
with $k=\sqrt{k_x^2+k_y^2}$, so that formally $\tilde{u}(k)=\frac{1}{\omega^2/c^2-k^2}\tilde{f}(k)$. Unlike the screened Poisson equation case, the denominator vanishes when $k^2=\omega^2/c^2$, making the inverse singular.

To regularize this singular behavior and ensure the correct causal solution, one introduces a small imaginary part
\begin{equation}
	\frac{1}{\omega^2/c^2-k^2}
	\stackrel{\epsilon\to 0}{=}
	\frac{1}{\omega^2/c^2-k^2+i\epsilon}.
\end{equation}

In this way, the inverse remains a well defined analytic function of $k$, so the same quantum-arithmetic framework can still be applied. We note that in the discretized problem, $\epsilon\sim\omega/c\Delta k$, set by the Fourier-mode spacing. Since quantum amplitudes are bounded by unity, the coefficient encoded by the quantum circuit is taken to be
\begin{equation}
	q(k)=\frac{\epsilon}{\omega^2/c^2-k^2+i\epsilon},
	\label{regularization}
\end{equation}
whose magnitude is at most one and reaches unity on resonance. The problem therefore reduces to evaluating this complex-valued function using quantum arithmetic. As in the example above, one can compute $q(k)$ and encode it into an auxiliary qubit amplitude, \eqref{aux_rotation}. The only essential modification is that complex-valued arithmetic necessitates the use of $2n_q$ qubits to represent both the real and imaginary parts.

\begin{figure}[t]
	\centering
	\includegraphics[width=1\linewidth]{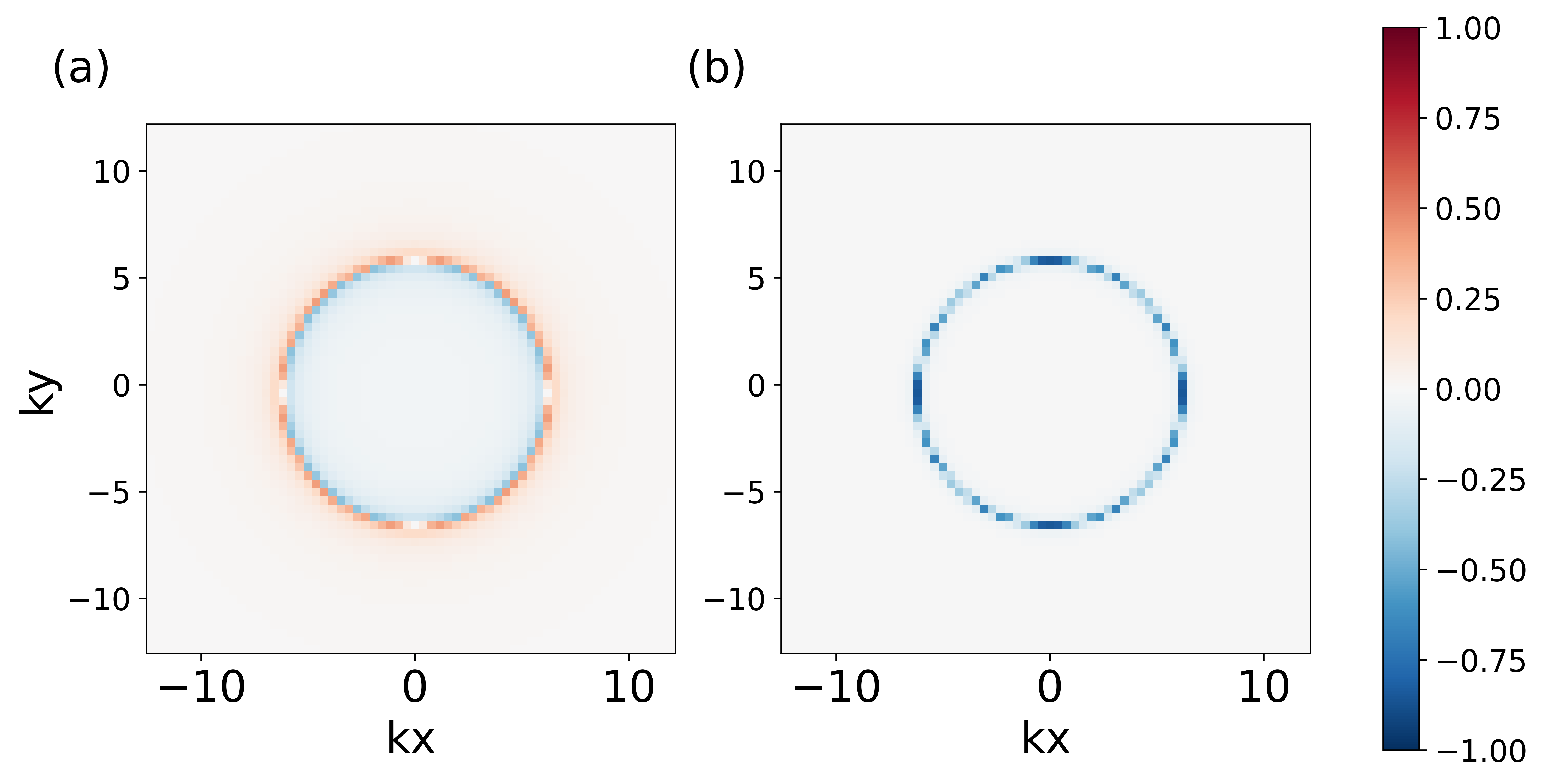}
	\caption{Illustration of the (a) real and (b) imaginary parts of Eq.~\eqref{regularization}, corresponding to the principal-value and delta-function contributions, respectively. The computation is performed on a discretized $64\times64$ grid over a square domain of side length $L=16\lambda$, where $\lambda$ denotes the wavelength. As seen, both terms are concentrated around the dispersion surface $k^2=\omega^2/c^2=2\pi/\lambda$. Here, $\epsilon=3\Delta k\omega/c$}
	\label{fig:PV_and_delta}
\end{figure}

One can gain further insight by decomposing \eqref{regularization} to its real and imaginary parts, using the identity
\begin{equation}
	\frac{1}{x+i\epsilon}
	\stackrel{\epsilon\to 0}{=}
	\mathrm{PV}\left(\frac{1}{x}\right)
	-i\pi\delta(x),
\end{equation}
where $\mathrm{PV}$ denotes the Cauchy principal value and $\delta$ is the Dirac delta function. One obtains
\begin{equation}
	\frac{1}{\omega^2/c^2-k^2+i\epsilon}
	\stackrel{\epsilon\to 0}{=}
	\mathrm{PV}\left(\frac{1}{\omega^2/c^2-k^2}\right)
	-i\pi\delta(\omega^2/c^2-k^2).
	\label{PV_and_delta}
\end{equation}
Thus, the inverse consists of a principal-value contribution together with a singular term supported on the resonant surface $k^2=\omega^2/c^2$, illustrated in Fig.~\ref{fig:PV_and_delta}.

However, this singular structure leads to a fundamental limitation. Since both terms vanish away from the dispersion surface (as seen in Fig.~\ref{fig:PV_and_delta}), only a small fraction of Fourier modes contribute significantly to the solution. In the two-dimensional setting, the dispersion surface is a circle (or a narrow ring) in Fourier space, while in three dimensions the corresponding surface is a sphere.

The success probability of the amplitude-encoding step therefore scales approximately as the number of modes near the resonance divided by total number of Fourier modes $N=2^n$.
Since the resonant surface has lower dimension than the full Fourier domain, this ratio decreases with system size. Equivalently, the number of repetitions required for successful post-selection grows exponentially with the number of qubits $n$. Thus, the singular structure of the wave equation prevents a straightforward extension of the arithmetic algorithm and motivates the search for a different strategy.

\section{The geometric quantum state preparation algorithm}
\label{geometric_algorithm}

\subsection{Algorithm description}
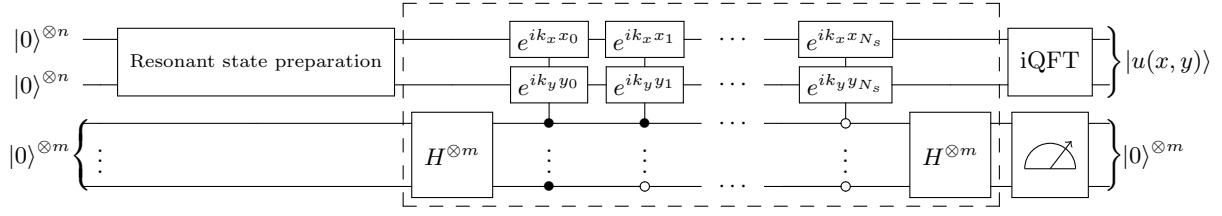
\begin{figure*}[t]
	\centering
\[
\Qcircuit @C=0.7em @R=0.38em {
	& & & \rule{0pt}{0.6em} & & & & & & & & & & 
	\\
	\lstick{\ket{0}^{\otimes n}}
	& \qw
	& \multigate{1}{\text{\scriptsize Resonant state preparation}}
	& \qw
	& \gate{e^{i k_x x_0}}\qwx[1]
	& \gate{e^{i k_x x_1}}\qwx[1]
	& \qw
	& \push{\cdots}
	&
	& \qw
	& \gate{e^{i k_x x_{N_s}}}\qwx[1]
	& \qw
	& \multigate{1}{\mathrm{iQFT}}
	& \qw
	\\
	\lstick{\ket{0}^{\otimes n}}
	& \qw
	& \ghost{\text{\scriptsize Resonant state preparation}}
	& \qw
	& \gate{e^{i k_y y_0}}
	& \gate{e^{i k_y y_1}}
	& \qw
	& \push{\cdots}
	&
	& \qw
	& \gate{e^{i k_y y_{N_s}}}
	& \qw
	& \ghost{\mathrm{iQFT}}
	& \qw
	\\
	& \qw
	& \qw
	& \multigate{2}{H^{\otimes m}}
	& \ctrl{-1}
	& \ctrl{-1}
	& \qw
	& \push{\cdots}
	&
	& \qw
	& \ctrlo{-1}
	& \multigate{2}{H^{\otimes m}}
	& \multigate{2}{\metersymbol}
	& \qw
	\\
	\lstick{\ket{0}^{\otimes m}}
	& \raisebox{0.15em}{\ensuremath{\vdots}}
	&
	& \nghost{H^{\otimes m}}
	& \raisebox{0.15em}{\ensuremath{\vdots}}
	& \raisebox{0.15em}{\ensuremath{\vdots}}
	&
	&
	&
	&
	& \raisebox{0.15em}{\ensuremath{\vdots}}
	& \nghost{H^{\otimes m}}
	& \nghost{\metersymbol}
	&
	\\
	& \qw
	& \qw
	& \ghost{H^{\otimes m}}
	& \control\qw
	& \ctrlo{0}\qw
	& \qw
	& \push{\cdots}
	&
	& \qw
	& \ctrlo{0}\qw
	& \ghost{H^{\otimes m}}
	& \ghost{\metersymbol}
	& \qw
	\inputgroupv{4}{6}{0.20em}{-0.15em}{}
	\outputgroupv{2}{3}{14}{0.18em}{2.4em}{0em}
	{\ket{u(x,y)}}
	\outputgroupv{4}{6}{14}{0.18em}{1.8em}{0em}
	{\ket{0}^{\otimes m}}
	\gategroup{1}{4}{6}{12}{0.7em}{--}
}
\]

\caption{Quantum circuit for the geometric quantum state preparation algorithm. The momentum registers are first initialized in the resonant state by the state-preparation procedure described in Appendix~\ref{appendix:ring_preparation}. The source-dependent phase encoding, highlighted by the dashed box, is then implemented by controlled phase gates that apply the phases $e^{i(k_xx_j+k_yy_j)}$. Finally, after unpreparing the source register and post-selecting it onto $\ket{0}^{\otimes m}$, inverse quantum Fourier transform (iQFT) is applied to the momentum registers, yielding the solution state $\ket{u(x,y)}$.}

\label{fig:geometric_algorithm}
\end{figure*}

We overcome this difficulty by introducing a new approach - \emph{geometric quantum state preparation}. Rather than preparing the full Fourier space and then post-selecting onto the resonant manifold, the algorithm prepares only the resonant components from the outset by constructing a quantum state supported on the resonant surface $k^2 \approx \frac{\omega^2}{c^2}$. In the notation of the previous section, this method directly prepares the action of $UA^{-1}$, after which applying $U^{\dagger}$ is the only remaining step. In this way, the exponentially small success probability associated with projecting from the full domain onto the resonant manifold is avoided entirely. 

To illustrate the algorithm, consider first the simple case of a single point source located at $(x_s,y_s)$. The goal is to prepare the Fourier state
\begin{equation}
	\frac{1}{\sqrt{\mathcal{N}r}}
	\sum_{k_x^2+k_y^2\approx \omega^2/c^2}
	q(k)\,
	e^{-i(k_xx_s+k_yy_s)}
	\ket{k_x,k_y},
	\label{state_final}
\end{equation}
where $\mathcal{N}r=\sum_{k_x^2+k_y^2\approx \omega^2/c^2}\left|q(k)\right|^2$ is the normalization factor associated with the resonant Fourier modes. The coefficient $q(k)$ is given by Eq.~\eqref{regularization}, while the phase factor accounts for the source position.

This construction extends naturally to multiple point sources. For sources located at $(x_j,y_j)$ with complex weights $w_j$, the target state becomes
\begin{equation}
	\frac{1}{\sqrt{\mathcal N}}
	\sum_{k_x^2+k_y^2\approx \omega^2/c^2}
	\left(
	\sum_j
	w_j q(k)
	e^{-i(k_xx_j+k_yy_j)}
	\right)
	\ket{k_x,k_y},
	\label{state_final_w}
\end{equation}
where $\mathcal N$ is the normalization constant. Thus, the phase contributions from all sources are coherently combined according to their corresponding weights.

The implementation naturally decomposes into two steps. In the first step, a resonant state is prepared by constructing an equal superposition over the Fourier modes contained within a narrow ring surrounding the resonant manifold and assigning to each momentum component the corresponding amplitude $q(k)$. The ring width, $r$, determines the approximation accuracy, increasing it includes more Fourier components, providing a more accurate representation of the regularized inverse operator $q(k)$ at the expense of a lower post-selection success probability. In the second step, the Fourier phases associated with the source positions are applied, yielding the desired propagated source state. A circuit diagram is shown in Fig. \ref{fig:geometric_algorithm}.

The construction of the resonant ring begins by evaluating the dispersion relation that defines the inner boundary,
$
r_-=\omega/c-r/2.
$
The central ingredient is a quantum square-root algorithm~\cite{wang2020quantum}, which evaluates
$
k_y=\pm\sqrt{r_-^2-k_x^2},
$
and, starting from a uniform superposition over the $k_x$ register, generates the corresponding values in the $k_y$ register, thereby preparing the arc of the inner boundary satisfying $|k_x|\le r_-/\sqrt{2}$. The prepared arc is then coherently extended up to the outer radius
$
r_+=\omega/c+r/2,
$
producing a uniform superposition over the corresponding sector of the resonant ring. The remaining sectors are subsequently generated by exploiting the symmetry of the circle, yielding a uniform superposition over all momentum components contained within the resonant ring. This procedure is the quantum analogue of the Midpoint circle (Bresenham) algorithm~\cite{bresenham1977linear,van1984efficient}. Finally, quantum arithmetic is used to encode the coefficient $q(k)$ of Eq.~\eqref{regularization}. The implementation details of the resonant-state preparation, including its quantum arithmetic subroutines and the corresponding quantum circuits, are presented in Appendix~\ref{appendix:ring_preparation}.

By preparing the state directly on the resonant manifold, the exponential suppression associated with post-selection from the full Fourier space is avoided. The remaining post-selection only reflects the amplitude encoding of $q(k)$ within the resonant manifold, giving a success probability equal to the average value of $|q(k)|^2$ over the resonant ring
\begin{equation}
	P_{\mathrm{succ}}
	=
	\frac{\sum_{k\in\mathrm{ring}}|q(k)|^2}
	{\sum_{k\in\mathrm{ring}}1}.
\end{equation}
In the continuous limit and using $k=\omega/c+x$, such that
$\omega^2/c^2-k^2\simeq2(\omega/c)x$, we obtain (see Appendix \ref{appendix:ring_preparation})
\begin{equation}
	P_{\mathrm{succ}}
	\approx
	\frac{\epsilon}{(\omega/c)r}
	\arctan\!\left(\frac{(\omega/c)r}{\epsilon}\right).
	\label{P_arctan}
\end{equation}
Setting the ring width to span $n_r$ Fourier grid spacings, $r=n_r\Delta k$, and the regularization parameter to $n_\epsilon$ times the characteristic scale $(\omega/c)\Delta k$, i.e., $\epsilon=n_\epsilon(\omega/c)\Delta k$, yields
\begin{equation}
	P_{\mathrm{succ}}
	=
	\frac{n_\epsilon}{n_r}
	\arctan\!\left(\frac{n_r}{n_\epsilon}\right),
	\label{arctan}
\end{equation}
which is independent of the Fourier spacing $\Delta k$, and therefore of the domain size.

After preparing the resonant state, the next step is to incorporate the source-dependent Fourier phases. Suppose that the source consists of $N_s$ emitters located at positions $(x_j,y_j)$, with $j=0,\dots,N_s-1$. One introduces an additional register of $m=\lceil \log(N_s) \rceil$ qubits to label the different sources. A Hadamard transform prepares an equal superposition over the source labels, corresponding to equal source amplitudes. More generally, arbitrary source amplitudes can be encoded with a corresponding weighted superposition. Then, controlled phase operations apply the phase factor $e^{-i(k_xx_j+k_yy_j)}$ conditioned on the corresponding source state. Afterward, the source register is uncomputed by applying the inverse state-preparation circuit (for equal source weights, this is simply another Hadamard transform). Post-selecting the source register onto the all-zero state yields the desired Fourier-space state, which takes the form of Eq.~\eqref{state_final_w}. This step is illustrated in the dashed-highlighted box of Fig.~\ref{fig:geometric_algorithm}. Finally, applying the inverse QFT maps the system to spatial basis.

The overall success probability of the post-selection step depends on the number of sources. In particular, the success probability scales as
\begin{equation}
	P_{\mathrm{succ}} \sim \frac{1}{N_s}.
\end{equation}
In many relevant physical scenarios, the number of sources is small compared to the system size. In this regime, the success probability remains effectively constant, and the algorithm retains an exponential speedup with respect to the domain size.

The present formulation assumes periodic boundary conditions, the natural setting in which the governing operator is diagonalized by the Fourier transform. However, reflecting boundary conditions can be incorporated through the standard odd- or even-extension technique, in which the physical domain is embedded into a doubled periodic domain with the appropriate symmetry. The Fourier representation of the extended problem then automatically satisfies the original Dirichlet or Neumann boundary conditions. In the quantum setting, this extension requires only one additional qubit per spatial dimension to represent the doubled coordinate range.

\subsection{Approximation accuracy and scaling}

\begin{figure}[t]
	\centering
	\includegraphics[width=\columnwidth]{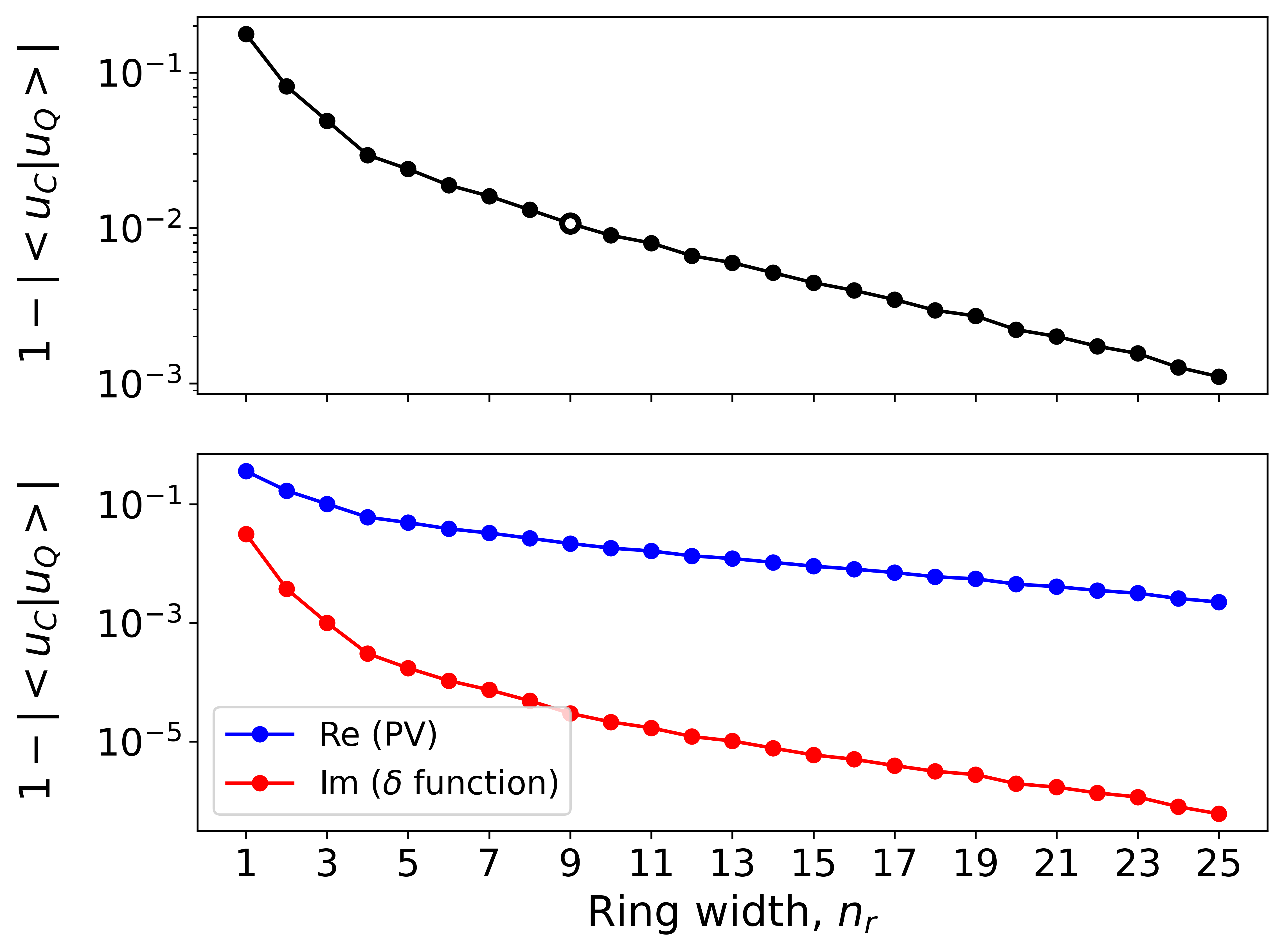}
	\caption{
		Accuracy of the resonant-ring approximation. (a) Total overlap error,
		$1-\left|\braket{u_C|u_Q}\right|$,
		between the normalized classical solution $\ket{u_C}$ and the state $\ket{u_Q}$ obtained from the resonant-ring approximation, as a function of the ring width, computed on a $64\times64$ spatial grid. The highlighted point corresponds to the simulation of Fig.~\ref{fig:results_demo}. (b) Error in the real and imaginary parts of the solution, corresponding to the real (principal-value) and imaginary (delta-function) parts of $q(k)$, respectively.
	}
	\label{fig:ring_error}
\end{figure}

\begin{figure}[t]
	\centering
	\includegraphics[width=\columnwidth]{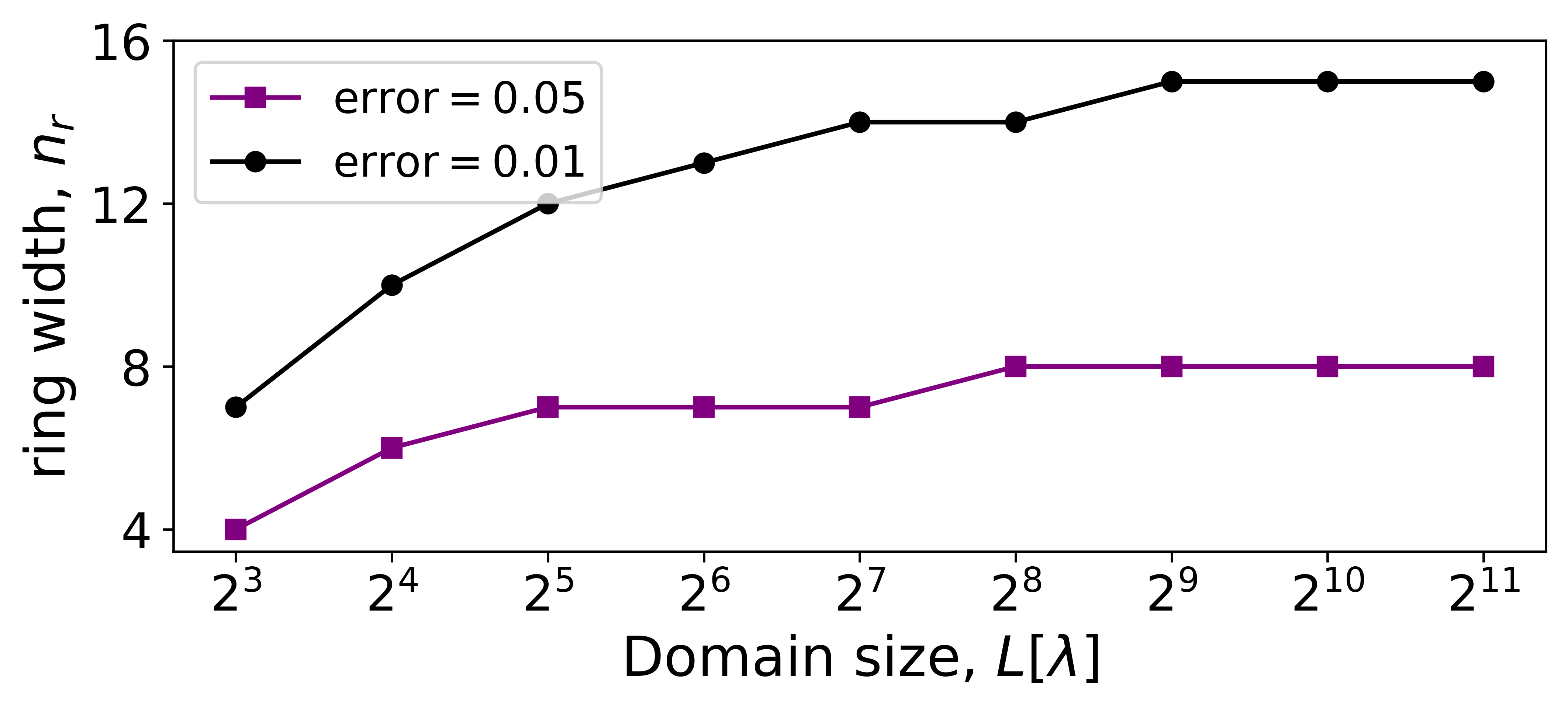}
	\caption{
		Minimum resonant-ring width required to achieve target approximation errors of $0.05$ and $0.01$ as a function of the computational domain size. The slight increase for the smallest domains is due to the coarse Fourier discretization. At higher resolution, the required ring width approaches a constant value, indicating independence from the domain size.
	}
	\label{fig:ring_scaling}
\end{figure}

The resonant ring introduces the only approximation in the algorithm, with its width controlling the trade-off between accuracy and post-selection probability. To quantify this approximation, Figs.~\ref{fig:ring_error} and \ref{fig:ring_scaling} compare the solution obtained from a classical emulation of the proposed quantum algorithm with the exact classical Fourier-space solution obtained numerically. In these simulations, the "quantum" solution is constructed exactly as in the proposed algorithm: rather than evaluating the contribution from the full Fourier domain, only the Fourier modes contained within the resonant ring are retained. The resulting state is then compared with the exact solution obtained from the complete Fourier-space calculation.

Fig.~\ref{fig:ring_error}(a) shows the overlap error,
\begin{equation}
	1-\left|\braket{u_C|u_Q}\right|,
\end{equation}
where $\ket{u_C}$ denotes the normalized classical solution and $\ket{u_Q}$ the state obtained from the resonant-ring approximation on a $64\times64$ spatial grid. All simulations in this subsection use $\omega/c=2\pi$ ($\lambda=1$), spatial discretization $h=\lambda/4$, and $n_{\epsilon}=3$. As expected, increasing the ring width incorporates additional Fourier components and yields a progressively more accurate representation of $q(k)$. Fig.~\ref{fig:ring_error}(b) separately shows the solution errors for the real and imaginary components, corresponding to the real (principal-value) and imaginary (delta-function) parts, respectively. The imaginary component converges rapidly because it is strongly localized around the resonant manifold and is therefore accurately captured by a relatively narrow ring. Consequently, the total approximation error is dominated by the more slowly converging real component.

Fig.~\ref{fig:ring_scaling} shows the minimum resonant-ring width required to achieve target approximation errors of $0.05$ and $0.01$ as the computational domain is increased over several orders of magnitude. The slight increase for the smallest domains is a finite-resolution effect caused by the coarse Fourier discretization. As the Fourier resolution improves, the required ring width approaches a constant value. These results indicate that maintaining a prescribed approximation accuracy requires only a constant ring width in Fourier space, independent of the computational domain size. Consequently, the resonant-ring approximation introduces only a constant computational overhead while preserving the asymptotic complexity of the algorithm.

\begin{figure}[t]
	\centering
	\includegraphics[width=1\linewidth]{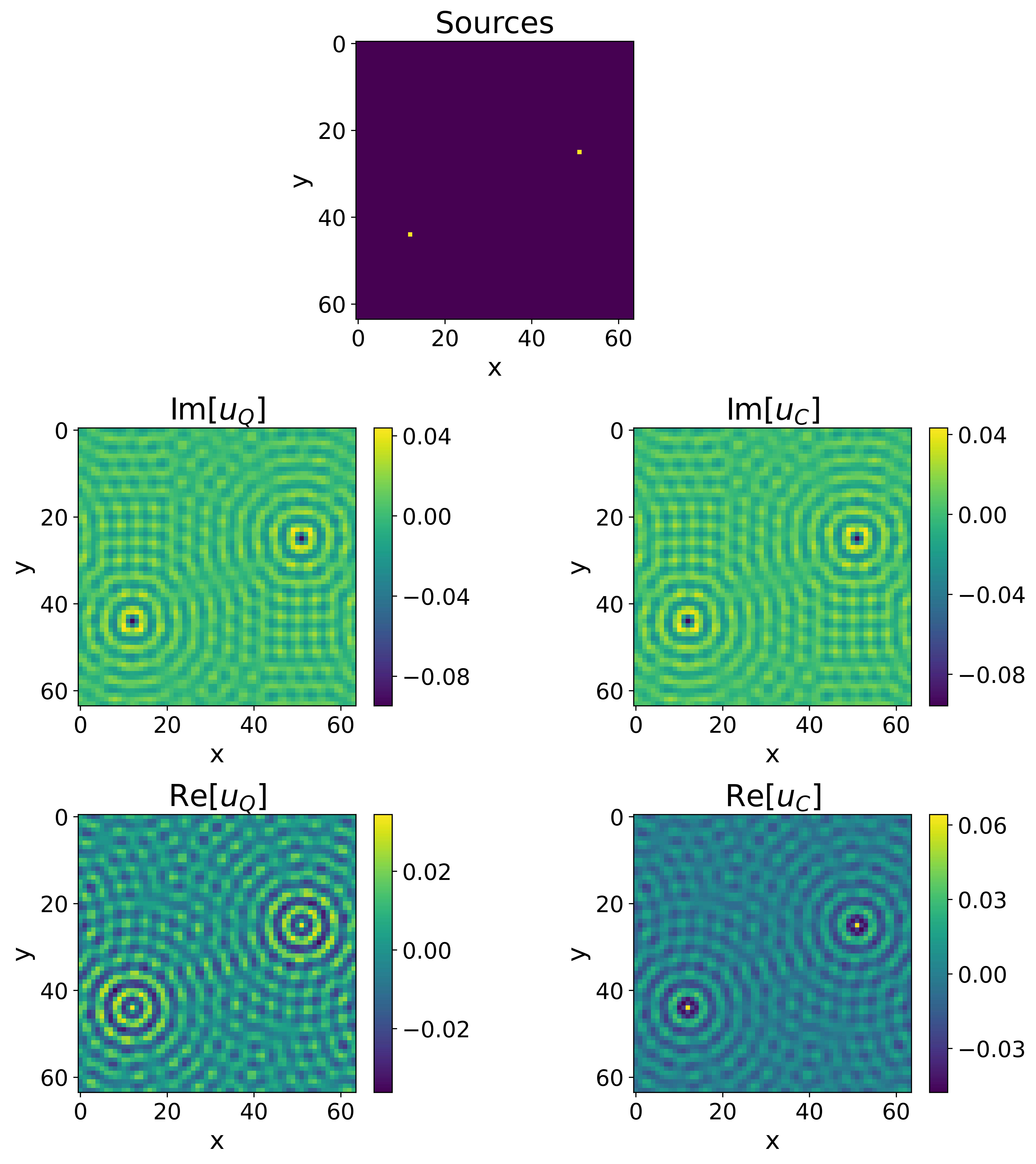}
	\caption{
	Demonstration of the geometric quantum state preparation algorithm for a system of size $64 \times 64$ (corresponding to $6+6$ data qubits). The top panel shows the source configuration consisting of two emitters. The middle and bottom rows compare the imaginary and real parts, respectively, of the wave field obtained from the classical emulation of the quantum algorithm, $u_Q$ (left), with the exact classical solution, $u_C$ (right). The overlap error between $u_Q$ and $u_C$ is $1-|\langle u_C|u_Q\rangle|=0.01$, corresponding to the highlighted point in Fig.~\ref{fig:ring_error}(a).
	}
	\label{fig:results_demo}
\end{figure}

Finally, we demonstrate the complete geometric quantum state preparation algorithm in Fig.~\ref{fig:results_demo}. Here, again, the spatial grid is $64\times64$, equivalent to $6+6$ qubits for the two spatial registers, consistent with the grid used in the preceding approximation-accuracy analysis. The resonant ring width is $r=9\Delta k$, $(n_r = 9)$, extending four Fourier grid points on either side of the resonant circle. The top panel shows the source configuration consisting of two localized emitters. The middle and bottom rows compare the imaginary and real parts, respectively, of the wave field obtained from the classical emulation of the quantum algorithm, $u_Q$, with the corresponding exact classical solution, $u_C$. The overlap error is $1-|\langle u_C|u_Q\rangle|=0.01$, corresponding to the highlighted point in Fig.~\ref{fig:ring_error}(a). The good agreement between $u_Q$ and $u_C$ demonstrates that the proposed algorithm accurately reproduces the expected wave propagation and interference patterns.

\subsection{Complexity analysis}

We now analyze the computational complexity of the proposed algorithm. To quantify its resource requirements, we consider the number of two-qubit gates (2QG), the total qubit count, and the expected number of repetitions due to post-selection.

\paragraph*{Gate count.}
We begin by analyzing the number of 2QG. As shown in Fig.~\ref{fig:geometric_algorithm}, the algorithm consists of resonant state preparation, source-dependent phase encoding, and an inverse quantum Fourier transform. The resonant state preparation, described in detail in Appendix~\ref{appendix:ring_preparation}, constitutes the dominant computational component, as it requires quantum arithmetic together with quantum comparator circuits. In particular, the evaluation of the square-root and reciprocal functions dominates the gate count. Efficient quantum implementations of these arithmetic operations have been presented in Ref.~\cite{wang2020quantum}, where both require $O(n^2)$ 2QG for an $n$-bit register. The remaining components of the algorithm consist primarily of controlled phase operations and a single inverse quantum Fourier transform, each requiring at most quadratic resources. Consequently, these contributions do not modify the overall asymptotic scaling, and the total gate complexity of the proposed algorithm is $O(n^2)$.

We also consider the $T$-gate count relevant for fault-tolerant implementations. Both the square-root and reciprocal circuits consist of $O(n)$ stages built from $n$-bit additions and subtractions~\cite{wang2020quantum}. Quantum adders and subtractors, have an $O(n)$ $T$-gate count~\cite{cuccaro2004new,paler2022realistic,remaud2024optimizing}. Consequently, the square-root and reciprocal circuits each require $O(n^2)$ $T$ gates. Comparator circuits can be constructed from the same addition and subtraction routines and likewise require $O(n)$ $T$ gates using at most one auxiliary qubit~\cite{cuccaro2004new,remaud2024optimizing}. The controlled sign-flip and controlled-SWAP operations contain only $O(n)$ constant-size controlled gates and therefore contribute at most $O(n)$ $T$ gates. At fixed synthesis precision, the Fourier-transform and phase-encoding operations contribute at most $O(n^2)$ $T$ gates~\cite{ross2016optimal}. Therefore, the overall $T$-gate count scales as $O(n^2)$.

We note that the above analysis is presented for the two-dimensional case. The three-dimensional extension is expected to exhibit the same asymptotic scaling, since it relies on the same quantum arithmetic subroutines, the square-root and reciprocal evaluations, which dominate the gate count and retain $O(n^2)$ complexity. The additional spatial dimension introduces only a constant-factor overhead associated with the extra coordinate register.

\paragraph*{Qubit count.}
We next consider the total number of qubits required by the algorithm. The dominant contribution again arises from the resonant state preparation described in Appendix~\ref{appendix:ring_preparation}. In particular, the quantum square-root circuit requires $5n$ qubits for an $n$-bit register~\cite{wang2020quantum}. In our two-dimensional construction, the same register width is sufficient, since the radicand $(\omega/c)^2-k_x^2-k_y^2$ remains bounded by $(\omega/c)^2$. Thus, the square-root operation requires $5n$ qubits for the $2n$-qubit spatial register. The remaining components require substantially fewer qubits. Besides the spatial registers of size $2n$, the algorithm employs an offset register of $n_\ell=\left\lceil\log_2\!\left(\sqrt{2}\,n_r\right)\right\rceil=\left\lceil \frac{1}{2}+\log_2 n_r\right\rceil$. qubits for the ring construction, two $n_q$-qubit registers for evaluating the real and imaginary parts of $q(k)$, an $m=\lceil \log(N_s) \rceil$-qubit source register encoding $N_s$ sources, and only a constant number of auxiliary qubits for the comparator, controlled sign-flip, and controlled-SWAP operations. Since we consider the regime in which the number of sources is small, $m\ll n$, the total number of qubits required by these additional registers is less than the $5n$ qubits already required by the square-root circuit. These qubits can therefore be allocated within the square-root workspace and reused after uncomputation. Consequently, the overall qubit requirement is $5n$. We note that in three dimensions, the same $5n$-qubit requirement is expected, since the additional spatial register does not increase the square-root register width.

\paragraph*{Number of repetitions.}
Finally, we consider the expected number of repetitions due to post-selection. As shown in Appendix~\ref{appendix:ring_preparation}, a resonant-state preparation of radius $\rho$ succeeds with probability
\begin{equation}
	P_{\rm res}
	\simeq
	\frac{\omega/c h}{16\,n_r}
	\frac{n_\epsilon}{n_r}
	\arctan\!\left(\frac{n_r}{n_\epsilon}\right),
\end{equation}
which is independent of the computational domain size. The only additional post-selection is associated with the source register, whose success probability is $1/N_s$. Therefore, the overall success probability is
\begin{equation}
	P_{\rm tot}=\frac{P_{\rm res}}{N_s}.
\end{equation}

As an example, the simulations presented in Fig.~\ref{fig:results_demo} use the parameters $h=\lambda/4$, corresponding to $\rho=\omega/c=2\pi/\lambda$, together with a resonant ring spanning $n_r=9$ Fourier grid spacings and a regularization width of $n_\epsilon=3$ Fourier grid spacings. These values give
\begin{equation}
	P_{\rm res}
	\simeq
	\frac{\pi}{864}
	\arctan(3)
	\approx
	4.54\times10^{-3},
\end{equation}
corresponding to an average of approximately $220$ repetitions for a single source, representing only a limited constant-factor overhead. Importantly, in contrast to generic QLSS approaches, the algorithmic complexity is no longer governed by the condition number.

\section{Conclusion}

We have presented a quantum algorithm for frequency-domain wave propagation problems, formulated through the Helmholtz equation. The central idea is to exploit the diagonalization of the Helmholtz operator in the Fourier basis and to prepare the resonant manifold directly, thereby eliminating the exponentially small success probability associated with projection from the full Fourier space. The resulting algorithm achieves a gate complexity of $O(n^2)$, which is polylogarithmic in the number of grid points, together with only a constant-factor repetition overhead for a fixed number of sources. This scaling could enable simulations far beyond classical reach; for example, a direct discretization of a three-dimensional wave field spanning tens of kilometers at centimeter-scale wavelengths requires $10^{18}$--$10^{21}$ grid points, whereas the quantum formulation requires only on the order of $100$ logical qubits. The present formulation assumes periodic boundary conditions, while Dirichlet or Neumann boundary conditions can be incorporated through odd or even extensions, requiring only one additional qubit per spatial dimension.

The proposed framework can be extended to a broad class of linear partial differential equations whose governing operators are diagonalized by spectral transforms, such as the Fourier transform, and whose spectral-domain solutions are supported on lower-dimensional manifolds determined by the corresponding dispersion relations, e.g., wave propagation in anisotropic or elastic media. In such cases, solving the differential equation is reduced to preparing the corresponding spectral manifold together with the associated spectral phases, thereby replacing matrix inversion with geometric state preparation. The framework can also accommodate more general source distributions beyond point emitters, particularly those with a simple representation in Fourier space. We anticipate that this paradigm will provide a foundation for the development of efficient quantum algorithms for a broader class of wave-propagation and spectral problems.

\section*{Acknowledgements}
We thank Dr. Andrea Caputo for his contribution to this work.

\bibliography{references}

@article{tezuka2026quantum,
  title={Quantum algorithm for electromagnetic field analysis},
  author={Tezuka, Hiroyuki and Sato, Yuki},
  journal={International Journal for Numerical Methods in Engineering},
  volume={127},
  number={10},
  pages={e70344},
  year={2026},
  publisher={Wiley Online Library}
}

@article{childs2021high,
  title={High-precision quantum algorithms for partial differential equations},
  author={Childs, Andrew M and Liu, Jin-Peng and Ostrander, Aaron},
  journal={Quantum},
  volume={5},
  pages={574},
  year={2021},
  publisher={Verein zur F{\"o}rderung des Open Access Publizierens in den Quantenwissenschaften}
}

@article{berry2014high,
  title={High-order quantum algorithm for solving linear differential equations},
  author={Berry, Dominic W},
  journal={Journal of Physics A: Mathematical and Theoretical},
  volume={47},
  number={10},
  pages={105301},
  year={2014},
  publisher={IOP Publishing}
}

@article{cao2013quantum,
  title={Quantum algorithm and circuit design solving the Poisson equation},
  author={Cao, Yudong and Papageorgiou, Anargyros and Petras, Iasonas and Traub, Joseph and Kais, Sabre},
  journal={New Journal of Physics},
  volume={15},
  number={1},
  pages={013021},
  year={2013},
  publisher={IOP Publishing}
}

@article{wang2020quantum,
  title={Quantum fast Poisson solver: the algorithm and complete and modular circuit design: S. Wang et al.},
  author={Wang, Shengbin and Wang, Zhimin and Li, Wendong and Fan, Lixin and Wei, Zhiqiang and Gu, Yongjian},
  journal={Quantum Information Processing},
  volume={19},
  number={6},
  pages={170},
  year={2020},
  publisher={Springer}
}

@article{yuan2023improved,
  title={An improved QFT-based quantum comparator and extended modular arithmetic using one ancilla qubit},
  author={Yuan, Yewei and Wang, Chao and Wang, Bei and Chen, Zhao-Yun and Dou, Meng-Han and Wu, Yu-Chun and Guo, Guo-Ping},
  journal={New Journal of Physics},
  volume={25},
  number={10},
  pages={103011},
  year={2023},
  publisher={IOP Publishing}
}

@article{wang2024quantum,
  title={Quantum Poisson solver without arithmetic},
  author={Wang, Shengbin and Wang, Zhimin and Cui, Guolong and Shi, Shangshang and Shang, Ruimin and Li, Jiaxin and Li, Wendong and Wei, Zhiqiang and Gu, Yongjian},
  journal={Intelligent Marine Technology and Systems},
  volume={2},
  number={1},
  pages={3},
  year={2024},
  publisher={Springer}
}

@article{shringi2026structure,
  title={Structure-Preserving Quantum Simulation of Wave Equations on a Trapped-Ion Processor},
  author={Shringi, Abhishek and Wu, Hsuan-Cheng and Shokry, Ahmed and Li, Xiantao and Kandemir, Mahmut Taylan},
  journal={arXiv preprint arXiv:2607.28499},
  year={2026}
}

@article{bagherimehrab2023fast,
  title={Fast quantum algorithm for differential equations},
  author={Bagherimehrab, Mohsen and Nakaji, Kouhei and Wiebe, Nathan and Brennen, Gavin K and Sanders, Barry C and Aspuru-Guzik, Al{\'a}n},
  journal={arXiv preprint arXiv:2306.11802},
  year={2023}
}

@article{tong2021fast,
  title={Fast inversion, preconditioned quantum linear system solvers, fast Green's-function computation, and fast evaluation of matrix functions},
  author={Tong, Yu and An, Dong and Wiebe, Nathan and Lin, Lin},
  journal={Physical Review A},
  volume={104},
  number={3},
  pages={032422},
  year={2021},
  publisher={APS}
}

@article{harrow2009quantum,
  title={Quantum algorithm for linear systems of equations},
  author={Harrow, Aram W and Hassidim, Avinatan and Lloyd, Seth},
  journal={Physical review letters},
  volume={103},
  number={15},
  pages={150502},
  year={2009},
  publisher={APS}
}

@article{childs2017quantum,
  title={Quantum algorithm for systems of linear equations with exponentially improved dependence on precision},
  author={Childs, Andrew M and Kothari, Robin and Somma, Rolando D},
  journal={SIAM Journal on Computing},
  volume={46},
  number={6},
  pages={1920--1950},
  year={2017},
  publisher={SIAM}
}

@article{morales2411quantum,
  title={Quantum linear system solvers: A survey of algorithms and applications (2025)},
  author={Morales, MES and Pira, L and Schleich, P and Koor, K and Costa, PCS and An, D and Aspuru-Guzik, A and Lin, L and Rebentrost, P and Berry, DW},
  journal={arXiv preprint arXiv:2411.02522}
}

@article{costa2022optimal,
  title={Optimal scaling quantum linear-systems solver via discrete adiabatic theorem},
  author={Costa, Pedro CS and An, Dong and Sanders, Yuval R and Su, Yuan and Babbush, Ryan and Berry, Dominic W},
  journal={PRX quantum},
  volume={3},
  number={4},
  pages={040303},
  year={2022},
  publisher={APS}
}

@article{gu2025quantum,
  title={Quantum simulation of Helmholtz equations via Schr{\"o}dingerization},
  author={Gu, Anjiao and Jin, Shi and Ma, Chuwen},
  journal={arXiv preprint arXiv:2507.23547},
  year={2025}
}

@article{paler2022realistic,
  title={On the realistic worst-case analysis of quantum arithmetic circuits},
  author={Paler, Alexandru and Oumarou, Oumarou and Basmadjian, Robert},
  journal={IEEE Transactions on Quantum Engineering},
  volume={3},
  pages={1--11},
  year={2022},
  publisher={IEEE}
}

@incollection{remaud2024optimizing,
  title={Optimizing T and CNOT gates in quantum ripple-carry adders and comparators},
  author={Remaud, Maxime},
  booktitle={Proceedings of Recent Advances in Quantum Computing and Technology},
  pages={56--61},
  year={2024}
}

@article{cuccaro2004new,
  title={A new quantum ripple-carry addition circuit},
  author={Cuccaro, Steven A and Draper, Thomas G and Kutin, Samuel A and Moulton, David Petrie},
  journal={arXiv preprint quant-ph/0410184},
  year={2004}
}

@article{ross2016optimal,
  title={Optimal ancilla-free Clifford+ T approximation of z-rotations.},
  author={Ross, Neil J and Selinger, Peter},
  journal={Quantum Inf. Comput.},
  volume={16},
  number={11\&12},
  pages={901--953},
  year={2016}
}

@article{costa2019quantum,
  title={Quantum algorithm for simulating the wave equation},
  author={Costa, Pedro CS and Jordan, Stephen and Ostrander, Aaron},
  journal={Physical Review A},
  volume={99},
  number={1},
  pages={012323},
  year={2019},
  publisher={APS}
}

@article{suau2021practical,
  title={Practical quantum computing: Solving the wave equation using a quantum approach},
  author={Suau, Adrien and Staffelbach, Gabriel and Calandra, Henri},
  journal={ACM Transactions on Quantum Computing},
  volume={2},
  number={1},
  pages={1--35},
  year={2021},
  publisher={ACM New York, NY, USA}
}

@article{sato2024hamiltonian,
  title={Hamiltonian simulation for hyperbolic partial differential equations by scalable quantum circuits},
  author={Sato, Yuki and Kondo, Ruho and Hamamura, Ikko and Onodera, Tamiya and Yamamoto, Naoki},
  journal={Physical Review Research},
  volume={6},
  number={3},
  pages={033246},
  year={2024},
  publisher={APS}
}

@article{babbush2023exponential,
  title={Exponential quantum speedup in simulating coupled classical oscillators},
  author={Babbush, Ryan and Berry, Dominic W and Kothari, Robin and Somma, Rolando D and Wiebe, Nathan},
  journal={Physical Review X},
  volume={13},
  number={4},
  pages={041041},
  year={2023},
  publisher={APS}
}

@article{luangsirapornchai2025practical,
  title={Practical quantum circuit implementation for simulating coupled classical oscillators},
  author={Luangsirapornchai, Natt and Sanglaor, Peeranat and Sornsaeng, Apimuk and Bressan, St{\'e}phane and Chotibut, Thiparat and Suksen, Kamonluk and Chongstitvatana, Prabhas},
  journal={IEEE Access},
  year={2025},
  publisher={IEEE}
}

@inproceedings{gilyen2019quantum,
  title={Quantum singular value transformation and beyond: exponential improvements for quantum matrix arithmetics},
  author={Gily{\'e}n, Andr{\'a}s and Su, Yuan and Low, Guang Hao and Wiebe, Nathan},
  booktitle={Proceedings of the 51st annual ACM SIGACT symposium on theory of computing},
  pages={193--204},
  year={2019}
}

@article{martyn2021grand,
  title={Grand unification of quantum algorithms},
  author={Martyn, John M and Rossi, Zane M and Tan, Andrew K and Chuang, Isaac L},
  journal={PRX quantum},
  volume={2},
  number={4},
  pages={040203},
  year={2021},
  publisher={APS}
}

@article{dong2024feedforward,
  title={Feedforward quantum singular value transformation},
  author={Dong, Yulong and An, Dong and Niu, Murphy Yuezhen},
  journal={arXiv preprint arXiv:2408.07803},
  year={2024}
}

@article{motlagh2024generalized,
  title={Generalized quantum signal processing},
  author={Motlagh, Danial and Wiebe, Nathan},
  journal={PRX Quantum},
  volume={5},
  number={2},
  pages={020368},
  year={2024},
  publisher={APS}
}

@article{chakraborty2018power,
  title={The power of block-encoded matrix powers: improved regression techniques via faster Hamiltonian simulation},
  author={Chakraborty, Shantanav and Gily{\'e}n, Andr{\'a}s and Jeffery, Stacey},
  journal={arXiv preprint arXiv:1804.01973},
  year={2018}
}

@article{clader2023quantum,
  title={Quantum resources required to block-encode a matrix of classical data},
  author={Clader, B David and Dalzell, Alexander M and Stamatopoulos, Nikitas and Salton, Grant and Berta, Mario and Zeng, William J},
  journal={IEEE Transactions on Quantum Engineering},
  volume={3},
  pages={1--23},
  year={2023},
  publisher={IEEE}
}

@article{sunderhauf2024block,
  title={Block-encoding structured matrices for data input in quantum computing},
  author={S{\"u}nderhauf, Christoph and Campbell, Earl and Camps, Joan},
  journal={Quantum},
  volume={8},
  pages={1226},
  year={2024},
  publisher={Verein zur F{\"o}rderung des Open Access Publizierens in den Quantenwissenschaften}
}

@misc{pyqsp,
  author       = {Isaac L. Chuang and collaborators},
  title        = {pyqsp: A Python Package for Quantum Signal Processing},
  year         = {2021},
  howpublished = {\url{https://github.com/ichuang/pyqsp}},
  note         = {Accessed: 2026-03-20}
}

@article{sunderhauf2025matrix,
  title={Matrix inversion polynomials for the quantum singular value transformation},
  author={S{\"u}nderhauf, Christoph and N{\'e}meth, Zal{\'a}n and Walayat, Adnaan and Patterson, Andrew and Berntson, Bjorn K},
  journal={arXiv preprint arXiv:2507.15537},
  year={2025}
}

@article{sanders2019black,
  title={Black-box quantum state preparation without arithmetic},
  author={Sanders, Yuval R and Low, Guang Hao and Scherer, Artur and Berry, Dominic W},
  journal={Physical review letters},
  volume={122},
  number={2},
  pages={020502},
  year={2019},
  publisher={APS}
}

@article{van1984efficient,
  title={An efficient ellipse-drawing algorithm},
  author={Van Aken, Jerry R},
  journal={IEEE Computer Graphics and Applications},
  volume={4},
  number={9},
  pages={24--35},
  year={1984},
  publisher={IEEE}
}

@article{bresenham1977linear,
  title={A linear algorithm for incremental digital display of circular arcs},
  author={Bresenham, Jack},
  journal={Communications of the ACM},
  volume={20},
  number={2},
  pages={100--106},
  year={1977},
  publisher={ACM New York, NY, USA}
}

\cleardoublepage
\onecolumngrid

\appendix

\section{Implementation of the resonant-state preparation}
\label{appendix:ring_preparation}

This appendix describes the implementation of the resonant-state preparation outlined in Sec.~\ref{geometric_algorithm}. We first present an efficient algorithm for preparing a circle state, corresponding to the limiting case of a ring with a single grid point in width. This construction then serves as the basis for the finite-width ring preparation described below.

\begin{figure*}[t]
	\centering
	\resizebox{0.88\textwidth}{!}{
		\Qcircuit @C=0.65em @R=0.6em {
			\lstick{\ket{0}^{\otimes n}_{k_x}}
			& \gate{H^{\otimes n}}
			& \gate{
				\begin{array}{c}
					\text{\scriptsize Comparator}\\[1mm]
					\text{\scriptsize $|k_x|\leq \rho/\sqrt{2}$}
			\end{array}}
			& \multigate{1}{
				\begin{array}{c}
					\text{\scriptsize Quantum square root}\\[1mm]
					\text{\scriptsize $k_y=\sqrt{\rho^{2}-k_x^{2}}$}
			\end{array}}
			& \qw
			& \qw
			& \qw
			& \qw
			& \qw
			& \qw
			& \qswap
			& \qw
			& \qw
			\gategroup{1}{2}{2}{4}{2.1em}{--}
			\\
			\lstick{\ket{0}^{\otimes n}_{k_y}}
			& \qw
			& \qw
			& \ghost{
				\begin{array}{c}
					\text{\scriptsize Quantum square root}\\[1mm]
					\text{\scriptsize $k_y=\sqrt{\rho^{2}-k_x^{2}}$}
			\end{array}}
			& \qw
			& \qw
			& \gate{\text{\scriptsize Sign flip}}\qwx[1]
			& \ctrl{1}
			& \qw
			& \qw
			& \qswap\qwx[-1]
			& \qw
			& \qw
			\\
			\lstick{\ket{0}_{a_s}}
			& \qw
			& \qw
			& \qw
			& \qw
			& \gate{H}
			& \ctrl{-1}
			& \targ
			& \qw
			& \qw
			& \qw
			& \qw
			& \qw
			\\
			\lstick{\ket{0}_{a_{\rm sw}}}
			& \qw
			& \qw
			& \qw
			& \qw
			& \qw
			& \qw
			& \qw
			& \qw
			& \gate{H}
			& \ctrl{-3}
			& \gate{H}
			& \meter
		}
	}
\caption{
	Quantum circuit implementing the quantum analogue of the Midpoint
	(Bresenham) circle algorithm. The dashed block prepares the
	$45^\circ$--$135^\circ$ arc by first generating a uniform
	superposition in the $k_x$ register, retaining the components
	satisfying $|k_x|\leq \rho/\sqrt{2}$, and computing
	$k_y=\sqrt{\rho^2-k_x^2}$. A controlled sign-flip operation,
	$\ket{k_y}\rightarrow\ket{-k_y}$, followed by a CNOT from the sign bit
	of $k_y$ to the auxiliary $a_s$, generates the opposite arc
	deterministically and resets $a_s$ to $\ket0$. The CNOT control drawn
	on the bundled $k_y$ wire acts only on its sign bit. Finally, a
	controlled-SWAP operation exchanges $k_x\leftrightarrow k_y$ to
	generate the remaining circle segments, with a post-selection
	success probability of approximately $1/2$. Arithmetic work registers
	and auxiliary qubits used internally by the comparator and
	square-root circuits are omitted for clarity.
}
	\label{fig:ring_boundary}
\end{figure*}
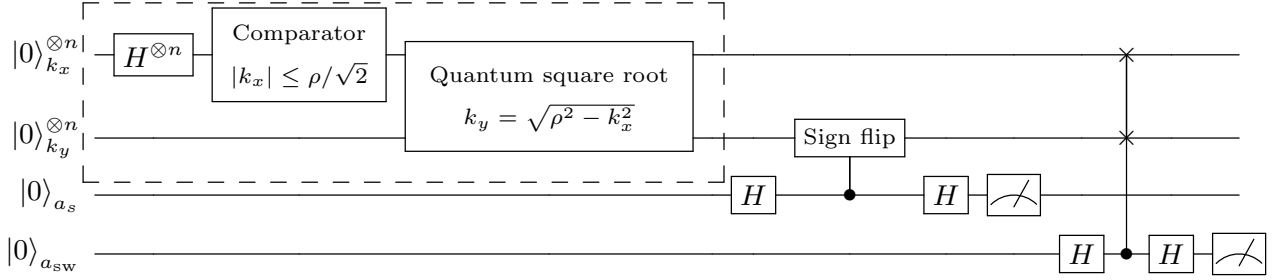

\par\medskip
\paragraph*{Circle state preparation.}

The circuit implementing the circle state preparation is shown in Fig.~\ref{fig:ring_boundary}, for a circle of radius $\rho$ (in the main text $\rho=\omega/c$). The construction begins by applying Hadamard gates to the $k_x$ register, producing an equal superposition of the signed numbers $k_x\in[-k_{\max},k_{\max}]$. A quantum comparator \cite{cuccaro2004new,yuan2023improved} then retains only those components satisfying
\begin{equation}
	|k_x|\leq \frac{\rho}{\sqrt2},
\end{equation}
corresponding to the two $45^\circ$ arcs adjacent to the positive and negative $k_x$ axes. The associated post-selection succeeds with probability
\begin{equation}
	P_{\rm comp}
	\simeq
	\frac{2\rho/\sqrt2}{2k_{\max}}
	=
	\frac{\rho}{\sqrt2\,k_{\max}}.
\end{equation}
For the Fourier discretization considered here, the momentum components span the interval $-\frac{\pi}{h}\leq k_x\leq \frac{\pi}{h}$, so that $k_{\max}=\pi/h$ and
\begin{equation}
	P_{\rm comp}
	\simeq
	\frac{\rho\,h}{\sqrt2\,\pi},
\end{equation}

We note that the grid spacing, $h=L/N$, is kept constant as the system size $L$ increases. Next, for each remaining value of $k_x$, a quantum square-root algorithm~\cite{wang2020quantum} evaluates
\begin{equation}
	k_y=\sqrt{\rho^2-k_x^2},
\end{equation}
and stores the positive solution in the $k_y$ register. 
To generate
the opposite arc coherently, an auxiliary qubit $a_s$ is initialized
in $\ket0$ and prepared in $(\ket0+\ket1)/\sqrt2$ using a Hadamard
gate. A controlled sign-flip operation, applied when $a_s=1$, produces
\begin{equation}
	\frac{
		\ket{k_x,k_y}\ket0_{a_s}
		+
		\ket{k_x,-k_y}\ket1_{a_s}
	}{\sqrt2},
	\qquad k_y>0.
\end{equation}
In two's-complement encoding, the sign bit $s_y$ of the momentum
register is $0$ in the positive-$k_y$ branch and $1$ in the
negative-$k_y$ branch. The auxiliary $a_s$ has the same value
in the corresponding branch. A CNOT controlled by $s_y$ and
targeting $a_s$ therefore leaves the auxiliary unchanged in
the first branch and flips it from $\ket1$ to $\ket0$ in the
second, yielding
\begin{equation}
	\frac{
		\ket{k_x,k_y}
		+
		\ket{k_x,-k_y}
	}{\sqrt2}\ket0_{a_s}.
\end{equation}
Thus, the opposite arc is generated deterministically, and the
auxiliary is reset and disentangled from the momentum registers
without measurement or post-selection.

The remaining circle segments are generated by exploiting the symmetry under the exchange $k_x\leftrightarrow k_y$. An auxiliary qubit prepared with a Hadamard gate controls a SWAP operation acting on the momentum registers. A second Hadamard and post-selection onto $\ket0$ project the registers onto the symmetric superposition
\begin{equation}
	\frac{\ket{k_x,k_y}+\ket{k_y,k_x}}{\sqrt2},
\end{equation}
with success probability of about $1/2$.

The resulting procedure is the quantum analogue of the Midpoint (Bresenham) circle algorithm~\cite{bresenham1977linear,van1984efficient}, where only one octant of the circle is computed explicitly while the remaining segments are generated using symmetry. Besides providing an efficient preparation of a single resonant circle, this construction serves as the geometric primitive underlying the finite-width ring preparation described below.
\par\medskip
\paragraph*{Finite-width ring preparation.}

\begin{figure*}[t]
	\centering
	\includegraphics[width=\textwidth]{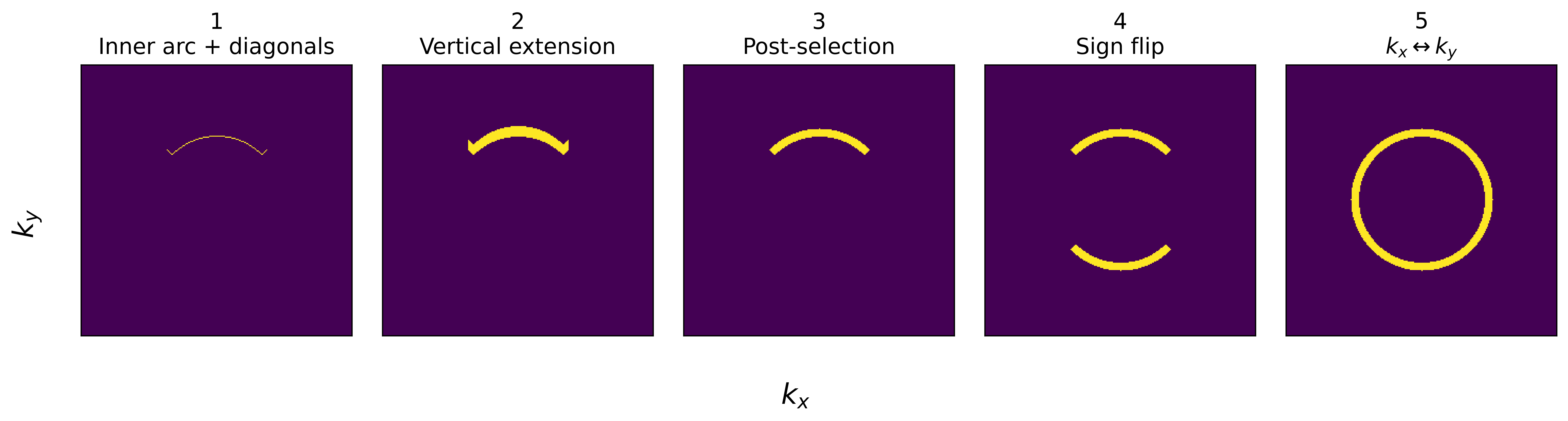}
	\caption{
		Construction of the finite-width resonant ring. 
		\textbf{(1)} The upper $45^\circ$--$135^\circ$ sector is initialized by combining the inner boundary prepared by the dashed block of Fig.~\ref{fig:ring_boundary} with the two diagonal segments extending from the inner to the outer radius.
		\textbf{(2)} A uniform superposition of vertical offsets is coherently added to every point of the initial curve, extending it across the ring width.
		\textbf{(3)} A quantum comparator removes the generated points lying outside the outer boundary, leaving the upper ring sector.
		\textbf{(4)} A controlled sign-flip operation,
		$\ket{k_x,k_y}\rightarrow\ket{k_x,-k_y}$, followed by a CNOT from
		the sign bit of $k_y$ to the controlling auxiliary, generates the
		corresponding lower sector deterministically and resets the auxiliary.
		\textbf{(5)} Finally, a controlled swap
		$k_x\leftrightarrow k_y$
		generates the remaining sectors, completing the resonant ring.
	}
	\label{fig:ring_construction}
\end{figure*}

We now describe the algorithm that constructs a uniform superposition over a ring of width $r$, with inner and outer radii $r_\pm=\rho\pm r/2$. As illustrated by the five steps in Fig.~\ref{fig:ring_construction}, the main idea is to prepare the curve bounding the upper $45^\circ$--$135^\circ$ sector, extend it vertically across the ring width, and then use sign and $x-y$ symmetries to complete the ring.

In step~1 of Fig.~\ref{fig:ring_construction}, an initial curve consisting of the inner $45^\circ$--$135^\circ$ arc and two diagonal segments is prepared. The role of the diagonal segments will be explained shortly. We first focus on the inner arc, which is prepared by applying the dashed block of Fig.~\ref{fig:ring_boundary} with radius $r_-$ and spans
\begin{equation}
	|k_x|\leq a,
	\qquad
	a=\frac{r_-}{\sqrt{2}}.
\end{equation}

The main idea of step~2 is to extend this arc vertically along the $k_y$ direction until it reaches the outer boundary. The largest required extension occurs at the endpoints $|k_x|=a$, where the vertical width of the ring is
\begin{equation}
	\Delta k_y^{\max}
	=
	\sqrt{r_+^2-a^2}-a
	\simeq
	\sqrt{2}r,
\end{equation}
with the approximation holding for $r\ll\rho$. We therefore extend the initial curve vertically by up to approximately $\sqrt{2}r$. To implement this extension, an auxiliary extension register of $n_\ell$ qubits is prepared in the uniform superposition
\begin{equation}
	\frac{1}{\sqrt{N_\ell}}
	\sum_{\ell=0}^{N_\ell-1}\ket{\ell},
	\qquad
	n_\ell=\left\lceil\log_2N_\ell\right\rceil .
\end{equation}
The number of extension values is chosen such that $N_\ell\Delta k\simeq\Delta k_y^{\max}$ or $N_\ell\simeq\frac{\sqrt{2}r}{\Delta k}$. A quantum adder then applies
\begin{equation}
	\ket{k_y}\ket{\ell}
	\mapsto
	\ket{k_y+\ell\Delta k}\ket{\ell},
\end{equation}
thereby generating the vertical extension. 

Extending the inner arc alone leaves two gaps because the arc is restricted to $|k_x|\leq a$, whereas the corresponding $45^\circ$--$135^\circ$ arc of the outer boundary reaches
\begin{equation}
	|k_x|=b,
	\qquad
	b=\frac{r_+}{\sqrt{2}}.
\end{equation}
If left uncorrected, the subsequent sign and $k_x\leftrightarrow k_y$ symmetry steps would leave four approximately triangular gaps around the diagonal directions of the final ring. To fill them, the initial curve prepared in step~1 also includes the diagonal segments $k_y=|k_x|$ for $a<|k_x|\leq b$, corresponding to the states $\ket{k_x,k_x}$ and $\ket{-k_x,k_x}$. To implement this correction, the $k_x$ register is prepared with
equal amplitudes over the full range $|k_x|\leq b$. The square-root operation $k_y=\sqrt{r_-^2-k_x^2}$ is conditioned on $|k_x|\leq a$, while for $a<|k_x|\leq b$ we set $k_y=|k_x|$. Thus, at the end of step~1, the $k_y$ register stores
\begin{equation}
	 k_y(k_x) = 
	 \begin{cases} 
	 \sqrt{r_-^2-k_x^2}, & |k_x|\leq a,\\[1mm] |k_x|, & a<|k_x|\leq b. 
	 \end{cases} 
\end{equation}

Step~3: Since the same extension of approximately $\sqrt{2}r$ is applied for every $k_x$, although the local vertical width of the ring varies with $k_x$, some generated points lie outside the outer boundary. Therefore, a quantum comparator retains only the states satisfying
\begin{equation}
	r_-^2
	\leq
	k_x^2+k_y^2
	\leq
	r_+^2,
	\label{the_inequality}
\end{equation}
while the remaining states are removed by post-selection. 

Let us now compute the success probability of this post-selection. Denoting by $N_x$ the number of initial $k_x$ values and by $N_{\rm in}(k_x)$ the number of points that satisfy \eqref{the_inequality}, the success probability is given by
\begin{equation}
	P_{\rm fill}
	=
	\frac{1}{N_xN_\ell}
	\sum_{k_x}N_{\rm in}(k_x).
\end{equation}
In the continuum limit, $N_{\rm in}(k_x)\Delta k$ approaches the required vertical extension,
\begin{equation}
	\Delta k_y(k_x)
	=
	\begin{cases}
		\sqrt{r_+^2-k_x^2}
		-
		\sqrt{r_-^2-k_x^2},
		& 0\leq |k_x|\leq a,\\[1mm]
		\sqrt{r_+^2-k_x^2}
		-
		|k_x|,
		& a<|k_x|\leq b.
	\end{cases}
\end{equation}
Using the symmetry under $k_x\rightarrow-k_x$, this gives
\begin{equation}
	P_{\rm fill}
	=
	\frac{
		\displaystyle
		\int_0^a
		\left[
		\sqrt{r_+^2-k_x^2}
		-
		\sqrt{r_-^2-k_x^2}
		\right]\,dk_x
		+
		\int_a^b
		\left[
		\sqrt{r_+^2-k_x^2}
		-
		k_x
		\right]\,dk_x
	}{
		\displaystyle
		b\,\Delta k_y^{\max}
	}.
\end{equation}
For $r\ll\rho$, the first integrand becomes
$\rho r/\sqrt{\rho^2-k_x^2}$. The diagonal interval has width
$b-a=r/\sqrt{2}$, so its contribution is of order $r^2$, whereas the arc contribution is of order $\rho r$. Consequently,
\begin{equation}
	P_{\rm fill}
	=
	\frac{
		\displaystyle
		\rho r
		\int_0^{1/\sqrt{2}}
		\frac{du}{\sqrt{1-u^2}}
		+
		\mathcal{O}(r^2)
	}{
		\displaystyle
		\rho r+\mathcal{O}(r^2)
	}
	=
	\frac{\pi}{4}
	+
	\mathcal{O}\!\left(\frac{r}{\rho}\right).
\end{equation}
Thus, the comparator post-selection succeeds with constant probability, approaching $\pi/4$ in the narrow-ring limit. 

After this post-selection, the inverse preparation of the offset register is applied, followed by projection onto $\ket{0}^{\otimes n_\ell}$. For a Hadamard-prepared offset register, this consists of a second layer of Hadamard gates. The conditional success probability is $1/N_\ell$. When the ring width is fixed as $r=n_r\Delta k$, one has $N_\ell\simeq\sqrt{2}\,n_r$, so this probability is also independent of the real-space domain size.

In step~4, the controlled sign flip generates the corresponding lower sector. In step~5, the controlled exchange $k_x\leftrightarrow k_y$ produces the remaining sectors and completes the ring. 

To complete step~5, we correct the amplitudes of diagonal points
satisfying $|k_x|=|k_y|$, which receive two identical contributions
from the post-selected swap and therefore have twice the amplitude
of off-diagonal points. An auxiliary qubit initialized in $\ket0$
is rotated to $\frac{1}{2}\ket0+\frac{\sqrt{3}}{2}\ket1$ conditioned on the diagonal points
and remains in $\ket0$ otherwise. Post-selecting
this auxiliary onto $\ket0$ halves the diagonal amplitudes while
leaving all other amplitudes unchanged, yielding a uniform ring
state. At fixed $n_r$, diagonal points form a vanishing fraction
of the ring as $L$ increases, so the success probability of this
correction approaches unity.

\par\medskip
\paragraph*{Amplitude encoding.}

The final part of the resonant state preparation assigns the coefficient $q(k)$ of Eq.~\eqref{regularization} to each momentum component. As illustrated in Fig.~\ref{fig:poisson_qft_inverse}, this is accomplished by evaluating $q(k)$ with reversible quantum arithmetic, storing its finite-precision representation in auxiliary registers, and using the result to control an amplitude-encoding operation. In our case $q(k)$ is complex, so separate arithmetic circuits are used to evaluate its real and imaginary parts, requiring two auxiliary registers of $n_q$ qubits each. The arithmetic circuits are subsequently uncomputed, disentangling the auxiliary registers from the momentum registers. Post-selecting the amplitude auxiliary then yields the desired resonant state
\begin{equation}
	\ket{\psi_{\rm res}}
	=
	\frac{1}{\mathcal N}
	\sum_{k\in\mathrm{ring}}
	q(k)\ket{k_x,k_y},
\end{equation}
where $\mathcal N$ is the normalization constant. Since the initial state is uniform over the ring, the success probability of this post-selection is given by the average of $|q(k)|^2$,
\begin{equation}
	P_{\mathrm{succ}}
	=
	\frac{\sum_{k\in\mathrm{ring}}|q(k)|^2}
	{\sum_{k\in\mathrm{ring}}1}.
\end{equation}

In the continuous limit, we write $k=\omega/c+x$, with $x\in[-r/2,r/2]$. In polar coordinates, $d^2k=kdkd\theta$, and since $|q(k)|^2$ depends only on the radial coordinate, the angular integral cancels between the numerator and denominator.
Near resonance, $\frac{\omega^2}{c^2}-k^2\simeq-2\frac{\omega}{c}x$, so that $|q(k)|^2$ is an even function of $x$. Consequently, over the symmetric interval $[-r/2,r/2]$, the contribution proportional to $x$ in the radial Jacobian $k=\omega/c+x$ vanishes, and the remaining factor $\omega/c$ cancels between numerator and denominator. We therefore obtain
\begin{equation}
	P_{\mathrm{succ}}
	\approx
	\frac{
		\displaystyle
		\int_{-r/2}^{r/2}
		\frac{\epsilon^2}
		{4(\omega/c)^2x^2+\epsilon^2}dx}
	{\displaystyle
		\int_{-r/2}^{r/2}dx}
	=
	\frac{\epsilon}{(\omega/c)r}
	\arctan\left(\frac{(\omega/c)r}{\epsilon}\right).
\end{equation}
This reproduces the scaling given in Eq.~\eqref{P_arctan} of the main text.

\par\medskip

\paragraph*{Overall success probability.}
Combining the post-selection steps, and using the
continuum and narrow-ring estimates derived above, the overall
success probability of the resonant-state preparation is
\begin{equation}
	\begin{aligned}
		P_{\rm res}
		&\simeq
		\frac{\rho h}{\sqrt{2}\pi}
		\cdot
		\frac{\pi}{4}
		\cdot
		\frac{1}{\sqrt{2}n_r}
		\cdot
		\frac{1}{2}
		\cdot
		\frac{n_\epsilon}{n_r}
		\arctan\!\left(\frac{n_r}{n_\epsilon}\right)
		\\
		&=
		\frac{\rho h}{16\,n_r}
		\frac{n_\epsilon}{n_r}
		\arctan\!\left(\frac{n_r}{n_\epsilon}\right).
	\end{aligned}
\end{equation}
The five factors correspond, respectively, to the initial arc
selection, the comparator removing points outside the ring, the
projection of the offset register onto its initial state, the
post-selection implementing the $k_x\leftrightarrow k_y$ symmetry,
and the amplitude encoding of $q(k)$
[Eq.~\eqref{arctan}]. The $k_y\leftrightarrow-k_y$ symmetry is
implemented deterministically and contributes no post-selection
factor. Since every remaining factor is independent of the computational
domain size, the expected number of preparation attempts remains
independent of the domain size.

\end{document}